# Personalized Metabolic Analysis of Diseases

Ali Cakmak and M. Hasan Celik

**Abstract**— The metabolic wiring of patient cells is altered drastically in many diseases, including cancer. Understanding the nature of such changes may pave the way for new therapeutic opportunities, as well as the development of personalized treatment strategies for patients. In this paper, we propose an algorithm, Metabolitics, which allows systems-level analysis of changes in the biochemical network of cells in disease states. It enables the study of a disease at both reaction- and pathway-level granularities for detailed and summarized view of disease etiology. Metabolitics employs flux variability analysis with an objective function which is dynamically built based on the biofluid metabolomics measurements in a personalized manner. Moreover, Metabolitics builds classification models for each disease to diagnose patients and predict their subgroups based on the computed metabolic network changes. We demonstrate the use of Metabolitics on three distinct diseases, namely, breast cancer, Crohn's disease, and colorectal cancer. Our results show that the constructed supervised learning models successfully diagnose patients by f1-score of over 90% on the average. Besides, in addition to the confirmation of previously reported breast cancer associated pathways, we discovered that Butanoate metabolism experiences significantly decreased activity, while Arginine and Proline Metabolism is subject to significant increase in activity, which have not been reported before. Metabolitics is made available as a Python package in pypi.

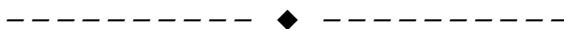

## 1 INTRODUCTION

THE phenotype of diseases has often reflections on the metabolism of patients [18], [42], [1], [11], [20]. Certain pathways may be boosted, while some others may experience activity decrease. Collectively, such changes may explain the etiology of a disease [18], [44]. In this paper, we propose an algorithm, Metabolitics, which quantifies the changes in activity levels of pathways (and their reactions) given concentration fold changes for a set of metabolites. It later employs the computed metabolic activity changes as features to build classification models that accurately diagnose patients.

A number of past studies (e.g., [13], [24], [56], [71], [77]) focused on pathway/reaction level analysis of high throughput biological data. The initial set of studies mainly employed pathway enrichment analysis ([56], [59], [68], [71], [72], [73], [74], [75], [76]). Briefly, these methods first identify significantly changing genes/metabolites in a given omics dataset. Then, the identified metabolites/genes are mapped to the pathways that they participate. Based on the number of changing metabolites/genes included in each pathway, statistical significance analysis is performed to identify those pathways that are significantly overrepresented with the measured genes or metabolites. Accordingly, each pathway is assigned a score based on the computed statistical test value. iOmicsPass [67] extends these approaches from node-level view to edge-level view in the context of the biological networks. More specifically, it does not consider the measured entities by themselves, but focuses on the direct interactions of the measured biological entities. To do this, it first computes co-expression scores for interactions based on their source and target entities' Z-scores. Next, using the interaction scores, it computes the significantly differing subnetworks. Then, on these computed subnetworks, pathway enrichment analysis is performed. XCMS Online [78] combines statistical significance-based filtering at both metabolite and pathway levels. The next set of methods ([13], [33], [50]) directly transfer the measured metabolite/gene changes to their corresponding pathways without any filtering. Then, statistical significance analysis is done on the pathways and their change levels to assign them a deregulation score. These methods consider each pathway as a collection of genes/metabolites, and ignore the interactions between these entities. An extension of the above set of methods also considers pathway topology ([12], [28], [52], [54], [69]). More specifically, some measured genes/metabolites are given more weights in the statistical significance analysis based on the centrality of each gene/metabolite in the pathway network topology. To quantify the centrality, different measures may be employed, such as betweenness, Eigen vector centrality (e.g., PageRank), neighborhood size, etc. [27].

The state of the art in this particular field is represented by Pathifier [13] and Paradigm [54]. Pathifier considers each pathway as a metabolite vector, matching the measured metabolites with the pathways that they participate. In this vector, each entry represents the concentration change measured for the metabolite corresponding to that vector row. Pathifier computes the principal curve that best fits the region where each individual is represented as a point in this high-dimensional space. On this curve, the mean distance from the points representing a person to the points representing healthy individuals is called pathway dysregulation score and shows how the pathway has changed in that individual.

Paradigm [54] is another successful approach in pathway-based analysis domain. It employs probabilistic

---

- A. Cakmak is with the Department of Computer Science and Eng., Istanbul Sehir University, Istanbul, Turkey. E-mail: alicakmak@sehir.edu.tr.
- M.H. Celik is with the Department of Computer Science and Eng., Istanbul Sehir University, Istanbul, Turkey. E-mail: hasancelik@std.sehir.edu.tr.

graph models to compute probability values for each pathway. A pathway is turned into what is called a *factor graph*. A factor graph contains a set of nodes representing different biological entities and their states. For instance, for each gene in a pathway, the factor graph contains the following nodes: gene, mRNA, protein, and active protein. The factor graph contains an edge for any kind of information flow or state change between these nodes. For instance, there is an edge from a gene to its mRNA representing transcription, an edge from mRNA to protein representing its translation, etc. Each node in a factor graph represents a variable which can only have one of following three discrete values: 1 (activation), 0 (normal), -1 (de-activation). The values of variables are learned from the provided omics data.

None of the above summarized pathway analysis methods considers the fact that pathways are part of a large biological network, and they interact with each other. The main novelty of the proposed method in this paper is that we perform the analysis on the whole pathway network in a holistic manner, rather than considering each pathway in an isolated manner. The main advantage of such an approach is that, for a given disease, it allows to identify those key player pathways for which there may be few or no associated gene/metabolite measurements in the analyzed omics data. Our method is not specific to the metabolomics domain. It may be easily extended to be used for the analysis of other types of omics data (e.g., mRNA) as well.

In brief, Metabolitics assigns a score for each pathway/reaction in a patient. This score represents how much the activity of the corresponding pathway/reaction differs from that of healthy individuals (in a negative or positive way). In order to achieve this, Metabolitics works on the whole network of metabolic pathways. It turns the analysis task into an optimization problem [38] where the objective is dynamically set to maximize the flux for increasing metabolites' reactions, and minimize the flux for decreasing metabolites' reactions in proportion to their fold changes. The metabolic network is assumed to be in steady state [35], that is, the amount of consumption and production is equal for all the metabolites. The steady state requirements are incorporated as constraints into the optimization problem. Then, the optimization problem is solved with linear programming [14]. Since the optimization problem is under-determined [38], there are usually multiple solutions. In order to accommodate multiple solutions with a single score, we employ flux variability analysis [31] to identify the lower and upper bound flux values for each pathway. The average lower and upper flux values of healthy individuals are considered as reference values. Then, for each pathway, Metabolitics compute how much the lower and upper flux values differ from the reference values in a given patient. The average of the changes in upper and lower flux bounds of a pathway is assigned as the "diff" score of the pathway. In this study, Metabolitics works on Recon consensus human metabolic network dataset [53].

Once "diff" scores are computed, Metabolitics builds supervised-learning models to automatically diagnose patients based on the computed metabolic changes. In partic-ular, each individual is represented as a vector of the computed personalized pathway diff scores.

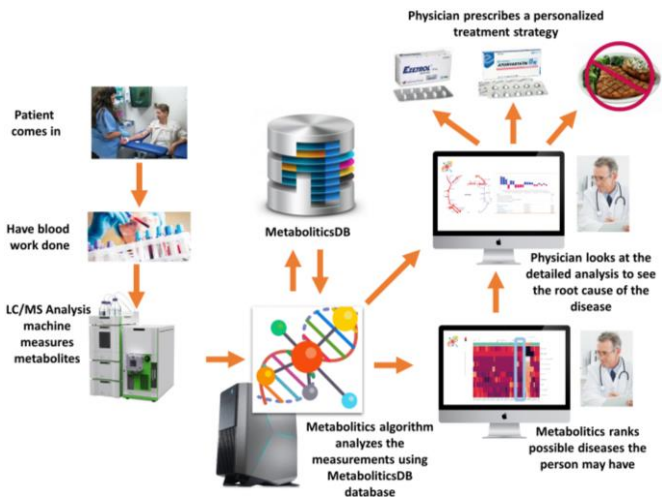

Fig. 1. Big Picture – A use case scenario for Metabolitics

Fig. 1 illustrates a real life use-case scenario for Metabolitics. More specifically, first, some blood sample is collected from an individual. Then, this sample is run through LC/MS process to identify the metabolites and their concentrations in the blood sample. Next, Metabolitics is run to identify the possible differentiations in the metabolic network of the individual. Then, using the pre-built classification models in its database, Metabolitics provides a physician with a set of likely diseases (if any) that the individual may have. At this point, if needed, the physician may further request more specialized tests before finalizing the diagnosis. Once the individual is diagnosed with a disease, the physician may look at the details of the computed metabolic changes in the individual. Accordingly, the root cause of the condition may be discovered, and an appropriate treatment strategy (e.g., a particular medicine, diet, etc.) may be planned by the physician in a personalized manner. As an example, suppose that an individual has high cholesterol condition. There may be at least three different underlying causes for high cholesterol [17]: (i) cholesterol may be excessively absorbed in intestines, (ii) too much cholesterol may be produced in the cells, or (iii) there may be a slowdown in the conversion of excess cholesterol to bile acids. For case (i), the physician may prescribe Ezetimibe (which inhibits the uptake of cholesterol through intestines). For case (ii), the physician may prescribe HMGCoA reductase inhibitor drugs which target endogenous cholesterol synthesis. On the other hand, for case (iii), there is no known medicine, and the only possible treatment strategy would be adopting a diet style which has no or low level of cholesterol.

In order to evaluate the Metabolitics algorithm, we apply it on breast cancer, Crohn's disease, and colorectal cancer. We demonstrate that Metabolitics (i) captures biologically relevant information, (ii) accurately diagnoses patients using machine learning, (iii) is robust to decrease in the amount of measurement data, and (iv) provides more coverage and robustness to data loss than the state of the art.

We have also implemented Metabolitics as a web-based tool, MetaboliticsDB (online at http://metabolitics.bi-odb.sehir.edu.tr), which is described in another work [7].

## 2 METHODS

In this section, we present the details of the Metabolitics algorithm. We use Recon2 dataset [53] as the metabolic network data. Recon2 is (genome-scale metabolic reconstruction) network model which includes 5324 metabolites and 7785 reactions in the humans. There are in total 100 pathways which are non-overlapping subgraphs of the network; thus, each has a unique set of reactions. Although, Recon2 used in this study for metabolite name matching, flux variability analysis, and feature extraction, Metabolitics is generic enough to adapt other network models.

In essence, the Metabolitics algorithm consists of the following steps.

1. Matching the names of the metabolites in the input metabolomics data to the metabolites in our database,
2. Dynamic creation of a linear programming model,
3. Flux variability analysis,
4. Calculation of reaction and pathway diff values,
5. Statistical significance analysis,
6. Feature extraction,
7. Feature elimination,
8. Building classification models for disease diagnosis

### 2.1 Matching Metabolite Names

It may not be possible to match all the metabolite names in a given metabolomics dataset to the metabolites in our database [41]. The main reason for this is that many metabolites do not have a standard common name used by all the researchers. Therefore, as a first step, one needs to find the equivalents of user-provided metabolites in our database. In this study, we compiled the set of all known alternative names for each metabolite by cross-integration of different metabolite data sources. More specifically, we combined the names of the metabolites that are included in two widely used, publicly available data sources, HMDB [58] and CheBI [23]. During this integration process, we considered two metabolite entries, one from HMDB and one from CheBI, as the same, if they share at least one synonym. Next, we search for each metabolite in our database in this combined name dataset, and transferred all known synonyms to our database for each matching record. Finally, we attempt to match the names in the metabolomics dataset given as input to the synonyms stored in the database.

### 2.2 Dynamically Constructing Linear Programming Models in a Personalized Manner

Metabolic analysis studies in the literature generally assume that metabolic networks are in steady state [30], [34]. The steady state hypothesis basically assumes that the total production of each metabolite in the metabolic network equals to its total consumption. This allows metabolic networks to be expressed mathematically as a set of linear equations. More specifically, a metabolic network can be expressed as a matrix in which the rows correspond to the metabolites and the columns correspond to the reactions in the network. In this matrix, a cell located at row $i$ and column $j$ contains the stoichiometry of metabolite $m_i$ in reaction $r_j$. If metabolite $m_i$ does not participate in reaction $r_j$, then 0 is placed in the corresponding cell. With this representation, flux balance analysis can be expressed as linear program under the steady state assumption as follows [38]:

$$maximize\ C^T \times V$$
$$subject\ to\ S \times V = 0\ and\ v_{lower} < V < v_{upper} \quad (1)$$

In the above equation, $S$ is the matrix representation of the metabolic network, $V$ is a vector of variables which represent reaction fluxes, $C$ is coefficient of reactions, $v_{lower}$ and $v_{upper}$ is the upper and lower boundary of the reactions' flux values. By solving the above linear program, the values of reaction flux variables in vector $V$ are determined. At this point, the steady state assumption is represented as constraints. In single cell organisms, the objective function is often set to maximize the fluxes of reactions that produce cell building blocks (amino acids, nucleotides, lipids, etc.) to amplify the cell biomass. However, for multi-cellular organisms, e.g., humans, there is no agreed-upon standard for the objective function structure. In this study, the objective function is dynamically set in a personalized manner. That is, the objective function (i.e., $C^T \times V$) is constructed as follows. $V$ includes flux variables for all reactions that produce at least one metabolite in the input metabolomics dataset. The corresponding reaction flux coefficients in $C$ are designated as specified in equation (2).

$$C_R = \sum_{m \in M_R} m_{fc}\ S_{m,R}\ /\ S_m^{total} \quad (2)$$

$$S_m^{total} = \sum_{r \in R} S_{m,r}$$

where
- $C_R$ is the coefficient of a reaction R's flux variable,
- $M_R$ is the set of metabolites that are produced by $R$,
- $m_{fc}$ is the measured fold change for metabolite m,
- $S_{m,R}$ is the stoichiometry of metabolite $m$ in $R$,
- $S_m^{total}$ is the total stoichiometry of m over all producer reactions of m in the metabolic network.

The objective function is heuristically built based on the intuition that, by incorporating $m_{fc}$ as a coefficient, our aim is to make sure that the higher the change in the measured concentration of a metabolite, the more effect it will have on the objective function value. In doing so, the total stoichiometry of the metabolite is used as a normalization term to prevent the popular (currency) metabolites from artificially dominating the objective function. This is because, in the metabolic network, there are some hub (currency) metabolites which contribute in many reactions. $H_2O$, $ATP$, and $NAD$ are some of the examples. Many studies [64] remove those metabolites from their analysis, but the removal of the currency metabolites may damage the stoichiometric balance of the metabolic network. Moreover, the removal of the currency metabolites is a strong assumption; therefore, should be avoided. In order to accommodate the above considerations, we keep the currency metabolites, and normalize each metabolite with its total stoichiometry. Moreover, the objective function is defined

---

**Flux variability analysis [63]:**

where solveFBA returns $C^T V$

**input:** FBA problem as $P$

**output:** minimum and maximum feasible flux for each reaction of FBA problem as $Rmax$ and $Rmin$

$solution$ = solveFBA($P$) with current objective
$P$ also subject to $C^T V = solution$

for each $r$ in $P.reactions$ do
    set maximization of $r$ as a new objective
    $Rmax_r$= solveFBA($P$)

    set minimization of $r$ as a new objective
    $Rmin_r$= solveFBA($P$)

---

completely based on the measured metabolite concentration changes which will be different for each person. This way of modeling allows our approach to produce person-specific results. Besides, metabolite concentrations need to be normalized as fold-changes since the scale of each metabolite is different than each other, and a change is only meaningful when compared to reference healthy value. That is,

$$m_i^{foldChange} = \log m_i^c - \log \mu_m^{healthy} \qquad (3)$$

where

- $m_i^c$ is the concentration measurement of metabolite $m$ in individual $i$.
- $m_i^{foldChange}$ is the concentration fold change of metabolite $m$ in individual $i$
- $\mu_m^{healthy}$ is the mean concentration of metabolite $m$ in all healthy individuals.

In order to compute the fold-changes as in eq. 3, we first compute the reference mean concentration of each metabolite based on the measurements in healthy individuals. Then, for all individuals (including healthy individuals and patients), we compute the metabolite fold changes relative to the above computed mean values. As a result, though usually smaller than patients, even healthy individuals are assigned metabolite fold changes. These fold changes are employed for each individual in the corresponding objective function as described above. We use this the objective function in flux variability analysis.

## 2.3 Flux Variability Analysis

Typically, the number of metabolic reactions is greater than the number of metabolites in a metabolic network [38]. Hence, the optimization problem presented in the previous section is underdetermined. Therefore, many alternative solutions (i.e., reaction flux value assignment) are usually possible. Each of these alternative solutions may allow the cells to achieve different goals. In order to cover all of these alternative solutions, flux variability analysis (FVA) [36] is employed. FVA allows to determine the minimum and maximum flux values for each reaction. In brief, FVA works as follows. First, the value of the objective function is determined by solving the optimization problem described in the previous section. Then, the value of the objective function, computed in the previous step, is added to the model as an additional constraint. For each reaction $R$ in the metabolic network, the objective function is set to maximize the flux of $R$, and the optimization problem is solved. The computed objective function value is stored as the upper boundary for the flux of $R$. Next, the objective function is set to minimize the flux of $R$, and the optimization problem is solved. The computed objective function value is stored as the lower boundary for the flux of $R$. This process is repeated for each reaction in the metabolic network.

## 2.4 Diff Value Computation

Pathways are defined by biologists as a set of closely related reactions that work together, often in a particular order, to achieve a common cellular goal (e.g., fatty acid synthesis). Analyzing the perturbations caused by diseases in metabolic networks in terms of changes in known pathways is a commonly used summarization method [13], [26]. In this study, we adopt a similar approach as well. To this end, for each pathway, we compute a "diff" value that represents the differentiation of the pathway activity for an individual in reference to healthy individuals. Pathway diff values are computed as follows.

First, the average lower and upper flux boundary values for each reaction is computed as described above based on the metabolomics data obtained from healthy individuals. These values are recorded as "reference" values. Then, during the analysis of a given set of metabolomics data of an individual, the difference between the lower and upper flux values obtained from this data set and the reference values are computed and recorded for each reaction as its "diff value". Then, for each pathway, its diff value is computed as the mean diff value of its reactions. This is similar to pathway-level aggregation enrichment methods [62] in the literature. As an alternative to computing the mean for all reactions, we also consider the mean of top-k reactions with the highest ANOVA score. A diff value may be negative or positive. Negative values indicate that the corresponding pathway/reaction is less active by the diff amount, while positive values indicate that the pathway experiences an activity increase with respect to healthy people.

## 2.5 Statistical Significance Analysis

It may be misleading to directly interpret a diff value solely based on its magnitude. In this study, we perform a statistical significance analysis to evaluate the possibility of random occurrence of these scores. To this end, ANOVA [9] is used to calculate the F- and p-values (corrected for multiple hypothesis testing using Benjamini-Hochberg), which indicate the statistical significance of pathway activity differences between patient and healthy groups. We prefer ANOVA because it is widely used in bioinformatics studies [61]. This way, the essential pathway changes that characterize a disease may be statistically determined. Such pathways may be focused to select the most appropriate treatment from existing therapies for personalized medicine, and for new drug target discovery.

## 2.6 Machine Learning-based Classification

Analysis results of metabolomics data from known (i.e., labeled) healthy individuals and patients are used as training data to build machine learning-based classification models. These trained models are then employed to predict whether a given metabolomics analysis results belong to a patient or a healthy person. In this context, each analysis result should first be transformed and represented as a numerical vector. To this end, one needs to first determine the structure and content of these numeric vectors.

**Feature Extraction:** Since our approach in this study is pathway-based, pathway diff values are used as features (i.e., entries in the numeric vector representation). There are 100 different pathways in the current version of the Recon data. Hence, in this study, the size of each vector representing an individual is 100.

**Feature Selection:** The employed features may be correlated among themselves, which may negatively affect the performance of the trained classification models. In order to alleviate this problem, as a preprocessing step, we eliminate the highly correlated features from the vector representation before building a classification model. We consider recursive feature elimination [22], feature importance based on feature weights, and ANOVA-based feature selection.

**Dimensionality Reduction:** We further perform dimensionality reduction. To this end, we consider the Principal Component Analysis [3], factor analysis [65], and Truncated SVD methods.

**Classification:** A classification model for a disease is constructed by using the vector representations of the metabolomics analysis results from healthy individuals and patients as training data. To this end, there are a number of alternative methods available in the literature. Among the most commonly used algorithms are Support Vector Machines [22], Decision Trees [29], and Logistic Regression [6]. In this study, we considered several classification methods, and employed Logistic Regression, as it provides the best or comparable performance. Repeated (10 times) stratified K-fold cross-validation is performed to test the performance of alternative machine learning methods. In each iteration, 8 folds are used for training, 1 fold is used for parameter tuning, and 1 fold is used for testing. In each repeat, the data is re-shuffled and re-splitted to k-fold. Reported results are average and standard deviation of the 100 folds (k=10, repeats=10).

## 3 EXPERIMENTAL EVALUATION

In this section, we present results from an experimental evaluation of Metabolitics on a breast cancer dataset [26] that includes around 160 metabolite measurements collected from 214 individuals' plasma samples (76 healthy individuals and 138 breast cancer patients). We also apply Metabolitics on Crohn's disease and colorectal cancer for further evaluation. We first present evaluation on breast cancer dataset in detail, and later discuss the results on Crohn's disease and colorectal cancer datasets.

The implementation of the presented algorithms are done in Python. ScikitLearn [39] library was used for machine learning models and statistical significance analysis. Cameo [14] library was used for FVA calculations.

### 3.1 Evaluation of Pipeline steps of Metabolitics

We benchmark each step of the pipeline in comparison to alternative approaches to justify our choice of algorithm in each pipeline step.

Pipeline steps until flux variable analysis are fixed in the Metabolitics. The remaining steps (such as feature selection, dimensionality reduction, etc.) are optional, and may be easily plugged-in (or out) using Metabolitics API. We compare the f1-score of each step that follows flux variable analysis with repeated stratified K-fold (10 fold and 10 repeat) cross-validation on breast cancer dataset. Results are reported in Figure 2. Higher classification performance achieved using reaction-level diff-scores. However, reaction level information alone is not interpretable due to a huge number of reactions (about 7500 of them). Thus, we aim to obtain pathway level score which achieves similar classification performance.

Averaging all reaction scores to obtain pathway scores slightly lowers scores because few number of reactions for each pathway is significantly altered and probably have important signal unlike the rest of the reactions whose change is insignificant. Thus, we choose top-k significant reactions (k=100) based on their ANOVA score, and then compute the average pathway diff score over top-k reactions. Tuning the number of those top-k reactions does not make much difference; thus, we use k=100 for this study. Reaction-level feature selection slightly decrease the performance. However, averaging to pathway-level slightly improve the performance of classification back to the same level. Feature selection at pathway-level contributes to the pathway-level performance, as it better represents pathway alteration characteristics. We have considered recursive feature elimination (89.8 ± 4.8) [22] and feature importance based on feature weights (89.4 ± 4.5) in logistics regression as alternatives to ANOVA. However, none of them provides better performance than ANOVA; thus, we employ and report ANOVA in feature selection.

### PCA

Metabolitics produces pathway scores as output. Metabolic pathways are beneficial for biological interpretation, but they may not describe the latent space of the metabolic network. For instance, some increased activity in one energy-related pathway is probably associated with a change in another energy-related pathway. This phenomenon is also seen in the breast cancer dataset. Figure 3 shows the variance of pathway score explained by PCA components. They clearly demonstrate that a few components of PCA explain most of the variance in pathway scores. The first component of the PCA explains ~47% of the variance, and the first ten components explain ~95% of the variance. Although there are 100 pathways in the metabolic network, 24 components of the PCA explains all the variance. Besides, Figure 4 visualizes the same dataset based on the first two components of the PCA, and there is a clear separation between healthy and breast cancer samples. Thus, we employ PCA for the classification task, and ANOVA-based feature importance scores of pathways for biological interpretation. Based on Figure 2, both methods are slightly better than raw pathway score. Moreover, there is no clear difference between PCA and reaction-level ANOVA-based feature selection. Moreover, we employ alternative dimension reduction methods such as factor analysis (82.4 ± 1) [65] and Truncated SVD (89.7 ± 4.4) [66], but they do not have better performance than PCA. Moreover, PCA is not a default part of the Metabolitics. It is only used in machine learning-related tasks due to faster training time. Pathway significance scores discussed in the paper are calculated without PCA using ANOVA, as this is a more likely scenario for most of the users who need enriched pathways for interpretation.

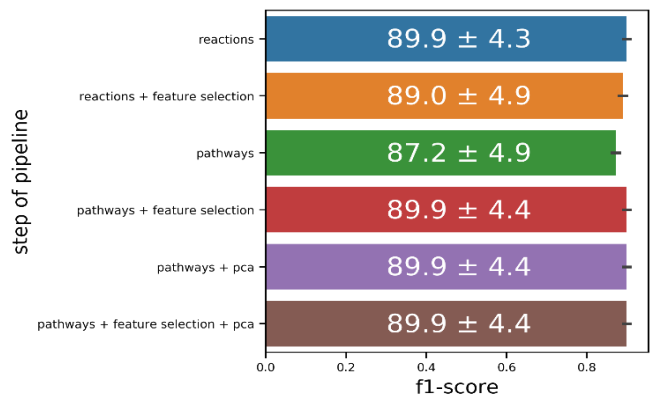

Fig. 2. Repeated stratified K-fold (10 fold and 10 repeat) cross-validation on breast cancer dataset (Feature selection is ANOVA-based).

**Classifiers**
We also consider alternative classifiers. Average and standard deviation of f1-scores for each classifier, which are calculated with repeated-stratified K-fold (10 fold and 10 repeat) cross-validation are presented in Figure 5. We tune hyper-parameters of each model, such as regularization parameter and different regularization metrics (L1 and L2) for logistics regression and SVM, gini and entropy as criterion metric and different tree depths for decision tree and random-forest, different numbers of estimators for ensemble methods (adaboost and random-forest), etc. Although most of the performance figures are within the error bars of the alternative methods, logistic regression and linear-svm show higher performance than others. In this study, a simpler model such as logistic regression is employed due to two main reasons. Firstly, available data is relatively small (214 breast cancer, 40 Crohn's disease, 114 Colorectal Cancer samples). Thus, training more recent complex models, such as deep neural networks, will likely overfit the data and not generalize. Moreover, we demonstrate that the performance of classification is not due to the learning capability or the complexity of the machine learning model, but the proposed subsystem analysis.

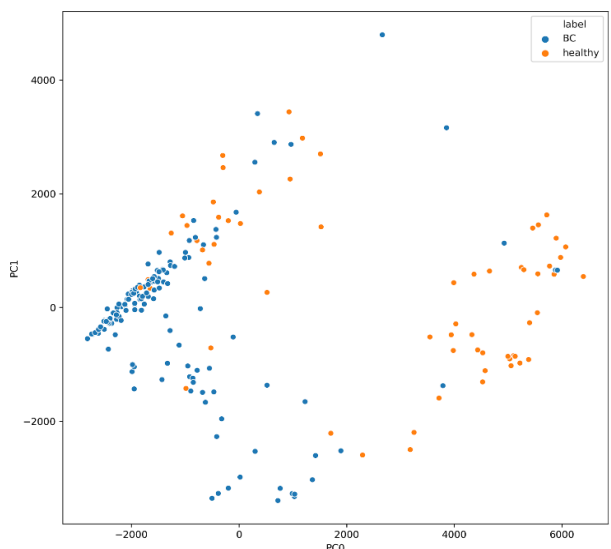

Fig. 4. The separation healthy and breast cancer samples with two PCA components

### 3.2 Metabolitics Captures Biologically Relevant Information in Breast Cancer
Table 1 lists the top-10 significantly changing pathways in breast cancer patients with their diff scores. Negative scores indicate that the average pathway flux decreases, and positive scores mean that the average pathway flux increases, in comparison to the healthy individuals. Fig. 6

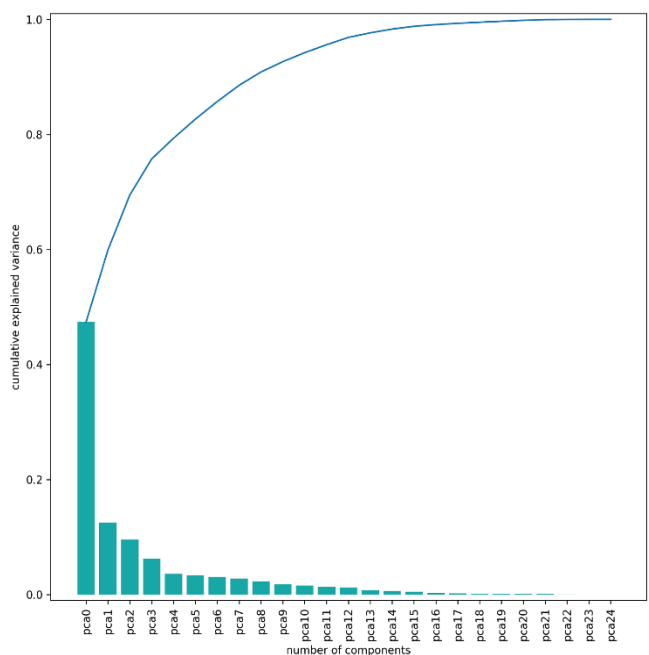

Fig. 3. The variance of pathway score explained by PCA components

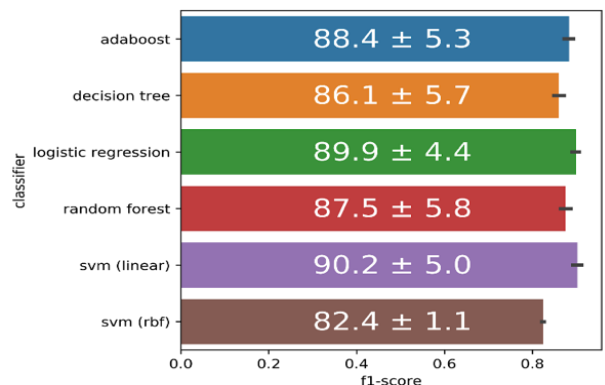

Fig. 5. f1-scores for different classification algorithms

shows the heatmap of the pathway diff values.

In Alanine and Aspartate Metabolism pathway, the reaction that experiences most flux increase is *asparagine synthase* (encoded by *ASNS*). This reaction converts *glutamine* into *glutamate* and *asparagine*. *Glutamate* is a precursor that leads to the biosynthesis of other amino acids which are needed by breast cancer tumor cells to proliferate [57].

The association of Arginine and Proline Metabolism with breast cancer is reported for the first time in this study. De Ingeniis et al. [10] suggested that *NADP+*, which is produced during the synthesis of *proline* from *arginine*, may be directed to *PP-ribose-P* synthesis. *PP-ribose-P* is later metabolized in nucleotide synthesis. Breast cancer tumors may also be using the same mechanism to enhance nucleotide synthesis.

Methionine and cysteine metabolism has the highest rate of increase in reaction *NADPH: CoA-glutathione oxidoreductase* (EC: 1.8.1.10). Breast cancer cells often experience high oxidative stress [4]. Through the conversion of *glutathione* (*GSH*) to its oxidized form, *glutathione disulfide* (*GSSH*), ROS are neutralized. Then, *glutathione reductase* converts *GSSH* back to *GSH* to process more ROS agents. Earlier studies (e.g., [40]) also report the increase of *GSH* and *glutathione reductase* activity in breast cancer tumors.

Similarly, *hypotaurine* also has a significant effect on the reactive oxidative stress states of cells [5], [19].

As related to the above phenomena, the *CoA* catabolism pathway exhibits a positive flux change. *Acetate* is an important input for lipid synthesis in breast cancer tumors [46]. More specifically, *Acetyl CoA* is first produced from *Acetate* (*ACOT12*) and then converted to *Malonyl CoA* (*ACC-alpha*) [47]. *Malonyl CoA* is the primary input metabolite in the first steps of fatty acid synthesis.

In Table 1, the activity of Fatty Acid Oxidation significantly reduces while the activity of the Fatty Acid Synthesis (not among top-10 pathways) increases significantly (p-val < 0.01). Fatty acids are used for lipid synthesis, and lipids form the main structure of the cell membrane. In order to form new tumor cells, lipids have a vital role as a building block of cell membranes [2]. Therefore, fatty acid synthesis is increased in tumor cells, and fatty acid oxidation is reduced.

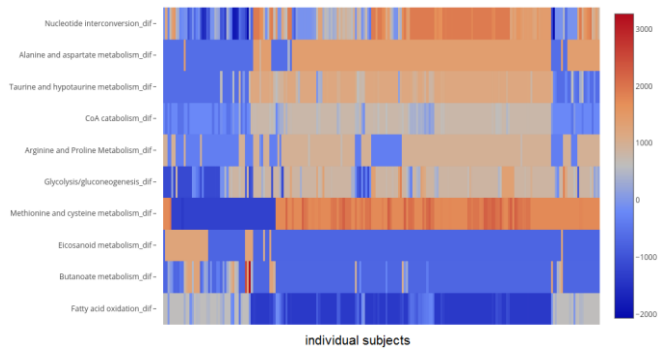

Fig. 6. Heatmap for the significantly changing top-10 pathways in breast cancer. Subjects are placed on x axis. Each cell is colored based on the diff values for the corresponding pathway.

*NTPs* and *dNTPs* form the main input metabolites of nucleotide synthesis. Nucleotide interconversion performs the transformations *(d) NMP ↔ (d) NDP ↔ (d) NTP* by transferring phosphate to produce the essential metabolites necessary for nucleotide metabolism. The increased activity of this pathway also indicates that the nucleotide synthesis in breast cancer tumor cells increases [48].

Eicosanoids are divided into 3 different subgroups: *prostanoids* (*COX*), *lipoxygenases* (*LOX*), and *ω-hydroxylases/epoxygenases*. *COX* and *LOX* eicosanoids have a supporting role in breast cancer development and metastasis [55]. *COXs* are more active in ER-positive breast cancer tumors, while *LOX's* are more effective in ER-negative tumors. In the ER-positive breast cancer type, tumor cells have estrogen sensitive receptors. Such tumors bind to the estrogen hormone, and estrogen accelerates the development and spread of these cells. COXs boost estrogen synthesis by increasing the activity of the rate limiting enzyme in estrogen synthesis, aromatase, and thus contribute to the development of ER-positive tumors [55]. Even though Eicosanoid metabolism seems to have a negative diff value in Table 1, when we examine diff values for the individual reactions of this pathway, we observe that the activity of reactions producing COX type eicosanoids has increased, and the flux value reactions producing other eicosanoid species has generally decreased. Since the number of reactions involved in the metabolism of non-COX eicosanoids was greater, the mean pathway differentiation value also comes out as negative. Based on the above observation, one may conclude that the majority of the breast cancer subjects included in the analyzed dataset belongs to the ER-positive subtype.

Butanoate is an important metabolite that has a therapeutic effect against breast cancer by directing the cells to apoptosis and stopping cell growth [51]. Therefore, the activity of Butanoate metabolism in patients is lower than healthy people.

Finally, the increase in Glycolysis points to what is known as Warburg Effect [43]. That is, tumor cells consume and convert glucose into lactate even though plenty of oxygen is available, and mitochondrial oxidation is possible.

Table 1. Significantly changing pathways in breast cancer with their diff scores computed by Metabolitics. The statistical significance values (i.e., F-val and p-val) are computed through ANOVA analysis based on computed diff values. The last column reports the measured metabolite counts for the reported pathways. Columns 5-6 provides the scores computed by Pathifier and Paradigm (discussed in Sec. 3.4)

| Pathway | F-val | p-val | diff | Pathifier | Paradigm | Met Cnt |
|---|---|---|---|---|---|---|
| Alanine and aspartate metabolism | 200 | 1.70E-31 | 1300 | 0.16 | 0.0012 | 7 |
| Arginine and Proline metabolism | 160 | 1.90E-26 | 850 | 0.46 | -0.0190 | 8 |
| Methionine and cysteine met. | 130 | 2.50E-23 | 170 | 0.17 | 0.0240 | 6 |
| Taurine and hypotaurine met. | 130 | 1.00E-22 | 970 | - | 0.1800 | 1 |
| CoA catabolism | 120 | 2.50E-22 | 610 | - | - | 0 |
| Fatty acid oxidation | 120 | 2.40E-21 | -1100 | 0.21 | - | 15 |
| Nucleotide interconversion | 110 | 2.70E-20 | 1300 | - | - | 1 |
| Eicosanoid met. | 80 | 6.30E-16 | -730 | 0.50 | - | 4 |
| Butanoate met. | 69 | 3.40E-14 | -670 | - | - | 0 |
| Glycolysis | 68 | 4.60E-14 | 770 | 0.11 | -0.0190 | 5 |
| Std. Deviation | | | 883.5 | 0.17 | 0.0800 | |

### 3.3 Metabolitics Accurately Diagnoses Breast

### Cancer Cases

We next demonstrate that Metabolitics allows to differentiate between healthy individuals and patients with high performance. In brief, we build a machine learning classification model based on the pathway diff score profile of each subject in our dataset as described in Methods section. In order to evaluate the validity of the classification model, we perform repeated stratified K-fold (10 fold and 10 repeat) [15]. We obtained the f1-score with the Logistic Regression [25] (C = 0.3e-6) algorithm as 89.9 ± 4.4.

### 3.4 Comparison with the State of the Art

We compare metabolitics with Pathifier [13] and Paradigm [54] in terms of (i) accuracy (AUC-ROC), (ii) coverage of the metabolism, and (iii) robustness to data loss.

**Accuracy**

Figure 7 presents AUC (Area under the Curve) ROC (Receiver Operating Characteristics) curve that is obtained with repeated stratified K-fold (10 folds and 10 repeats) cross validation. We use the default parameters of Pathifier and Paradigm, since they provide the best performance. Metabolitics has better AUC value than Paradigm [13]. Pathifier [53] seems to perform better than Metabolitics. However, these figures are misleading, as Pathifier and Paradigm overfit the data. More specifically, unlike Metabolitics, neither Paradigm nor Pathifier provides a programmatic API to split the train and test sets. That is, they learn and fit their model parameters on all the data. Paradigm scales the entire dataset. Thus, it also learns the mean of test dataset. Likewise, Pathifier employs the labels of the entire dataset to learn and extract pathway features. Thus, label information is also embedded into a generated feature which leads to the critical overfitting problem. Paradigm is proprietary software. Therefore, it is not available as open-source, and we had to use the binary executable of Paradigm to perform the comparison experiments in this study. Although Pathifier is available as open-source, its algorithm does not lend itself well for using in a cross-validation setting with different train and test datasets. The primary reason for this is that Pathifier is not originally designed with classification task in mind. Having said this, we have attempted to change the source code of the Pathifier, but we ended up with an algorithm that is significantly different than and may not be comparable to the original one. Thus, we aborted this option, and employed the original algorithms in our experiments despite the known overfitting problem.

To sum up, although Metabolitics cannot be compared with Paradigm and Pathifier in fair terms, it has comparable performance. Moreover, we provide a programmatic API to separate train-test data for our users to avoid the overfitting issue of state of the art algorithms.

**Coverage of the Metabolic Network**

In this section, we demonstrate that Metabolitics provides higher coverage of the metabolism than Paradigm and Pathifier. The coverage of the metabolic network is important to discover the complete set of metabolic changes caused by a disease.

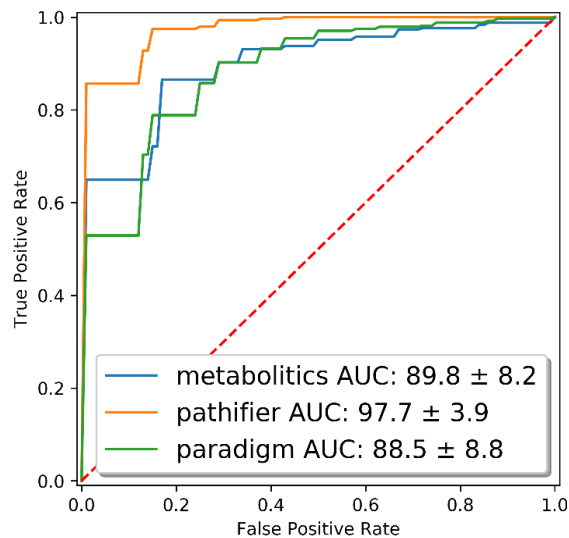

Fig. 7. AUC – ROC curve for Metabolitics, Pathifier, and Paradigm

According to Table 1, 4 of the 10 significant pathways reported by Metabolitics cannot be detected by Pathifier (i.e., no dysregulation score is computed). Similarly, Paradigm misses 5 of the top-10 pathways. The reason for this drawback of Pathifier and Paradigm is explained by the last column of Table 1. That is, Pathifier and Paradigm are only capable of evaluating and scoring pathways with at least a few metabolites included in the analyzed metabolomics data set. Therefore, it may miss many important key pathways for which no measured metabolite is available in the analyzed metabolomics dataset. On the other hand, since Metabolitics considers the metabolic network as a whole, and the pathways are interconnected in this network, it can produce analysis results about pathways for which no or few measurements are available.

**Robustness to Data Loss**

In this section, we compare the "robustness" of Metabolitics to that of Pathifier and Paradigm algorithms. We informally define the robustness as the resistance of an algorithm's performance to keep its initial value in the face of decrease in the number of measured metabolites. In other words, more robust algorithms experience lower performance loss, when the number of measured metabolites is less. In order to simulate the worst-case scenario, we first perform ANOVA to identify the most significantly differing metabolites between healthy individuals and patients. Next, we sort the metabolites by their F-values (as computed by ANOVA). Then, in an iterative way, we systematically remove the top-10 metabolites with the highest F-values from the metabolomics dataset, and then rebuild the classification models in each iteration on the reduced dataset. Fig. 8 charts the average f1-score of stratified K-fold (K=10) cross validation for Metabolitics, Pathifier, and Paradigm after each metabolite removal iteration. As shown, Metabolitics is more robust to decrease in measured data than Pathifier and Paradigm, since it provides insights on parts of the metabolic network with no metabolite measurements.

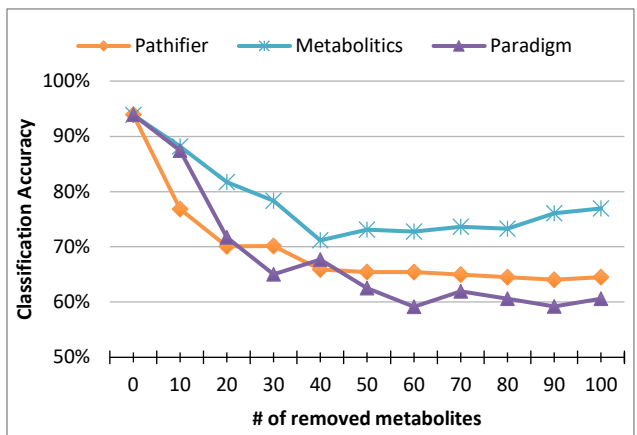

Fig. 8. Classification performance (average f1-score) of Metabolitics, Paradigm, and Pathifier with less measurements

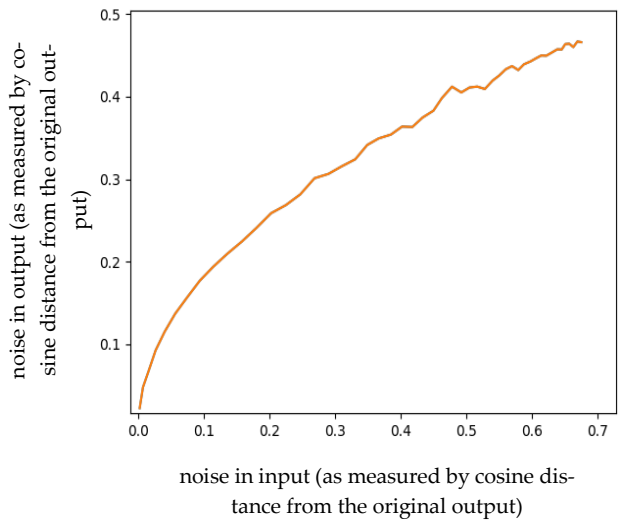

Fig. 9. The effect of noise in metabolomics data on the diff values

### 3.5 Pathway Significance-based Prediction

In this section, as an alternative method, we compute a pathway-wise significance that considers all positive and negative diffs in a pathway to determine whether the pathway is indeed up- or downregulated (similarly to gene set enrichment analysis, e.g., [60]). Fisher exact test is used to calculate pathway-wise significance. More specifically, average of healthy and patient samples are computed separately. As a result, for each pathway, we obtain two reaction vectors, one for healthy and one for cancer patients. Then, Fisher exact test is performed for each pathway based on these reaction vectors. Contingency table matrix has "healthy" and "patient" as columns, and the numbers of increasing, decreasing, and unchanged reactions as rows, respectively. The resulting significance values are employed as features in the classification task. The f1-score of this classification is 80%, which is lower than the original Metabolitics approach. The lower f1-score of the significance-based approach may be attributed to the following: (i) First, this approach only considers the direction of change in reaction fluxes (as either increasing or decreasing), and ignores the magnitude of change. (ii) Second, this approach just focuses on the number of reactions that increase or decrease, and does not consider the identity of these reactions, which leads to loss of information.

### 3.6 Stability Analysis

In this section, we further evaluate the stability of Metabolitics under systematically introduced noise in input data. The major goal of this analysis is to test whether Metabolitics amplifies the noise (introduced in the input metabolomics dataset) in its output diff scores. More specifically, in this analysis, some uniform noise between certain negative and positive range with increasing amount is added to input dataset to observe the effect of noise in the output of Metabolitics. Cosine distance is used to calculate the distance between the original and noise-added values in both input dataset and the corresponding Metabolitics-generated output. Randomly generated 100 samples are used as reference values, and their input and output values are averaged each time before the calculation of distance. This experiment is repeated 100 times for each noise interval and the results are averaged and presented Figure 9. All the experiments are run on E. Coli network to get results in a faster way, as a large numbers of runs are performed. Fig. 9 shows that the noise in the output grows slower than the rate of increase in the input noise. Thus, we conclude that Metabolitics does not amplify the noise or act unstable under noise.

Moreover, we further performed another experiment to measure the effect of noise in metabolomics measurements on f1-score in breast cancer dataset. In this experiment, uniform noises proportional to metabolite fold-changes are added with increasing amounts, and Metabolitics analysis is performed for each interval. Fig. 10 charts the change of f1-score as the noise amount increases. The graph shows that Metabolitics performs reasonably well under increasing noise amount as shown by the classification f1-score results.

### 3.7 Applying Metabolitics on other Diseases

In this section, to further validate the proposed methodology, we demonstrate the application of Metabolitics on two additional disease datasets, namely, Crohn's Disease [49] and Colorectal Cancer [45]. Fig. 11 presents the overall f1-score of Metabolitics in diagnosing different diseases.

**Crohn's Disease** (CD) is an autoimmune disorder that causes inflammation in the gastroenterological track of the patients. We run Metabolitics on a recent dataset [49] that is obtained from the serum measurements of 40 subjects (20 with CD and 20 healthy). With the Metabolitics assigned pathway diff scores as features, a Logistic Regression-based classifier (with C = 0.3e-6) differentiated subjects with CD from healthy ones with a mean f1-score of 88% (over 10-fold cross validation).

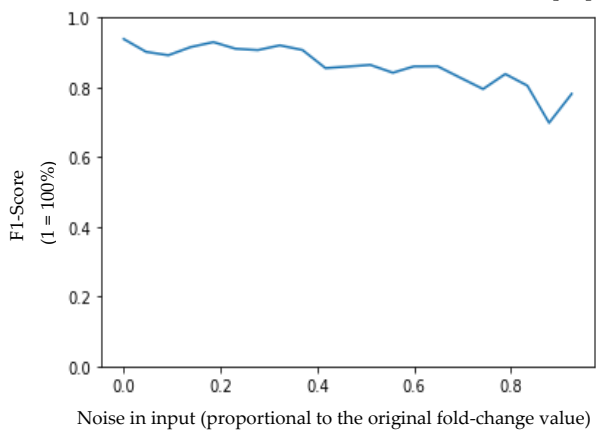

Fig. 10. The effect of noise in input metabolite measurements on the breast cancer diagnosis f1-score

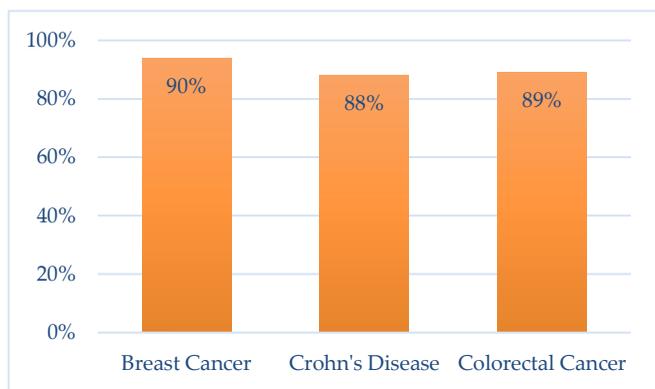

Fig. 11. Diagnosis f1-score of Metabolitics in different diseases

**Colorectal Cancer** (CRC) originates from the epithelial cells of colon or rectum, and it is one of the most common cancers. We run Metabolitics on a metabolomics dataset [45] that are obtained from Visceral Fat tissues of 55 subjects (49 CRC patients, 6 healthy control). On this dataset, Logistic regression achieves 89% f1-score on classifying CRC patients and healthy samples.

## 3.8 Sparsity Robustness Analysis

In this section, we perform a simulation study to analyze the effect of sparsity in terms of the coverage of metabolomics measurements of the metabolic network over the stability of Metabolitics analysis results. The major challenge in this experiment is the generation of metabolomics data that is as much realistic as possible, and covers the whole network. As a baseline approach, we ignored the network relationship between metabolites and assumed that all metabolites are independent of each other. More specifically, we generate a random metabolomics dataset which has full coverage of the network by assuming normal distribution with a mean of 0 and a standard deviation of 1. To perform the simulation, we systematically decreased the network coverage by 5% by randomly removing metabolites from the initial data set at each iteration. At each step, we ran Metabolitics on the current data set, and compute the correlation between the current results and that of full coverage dataset. The results are presented in

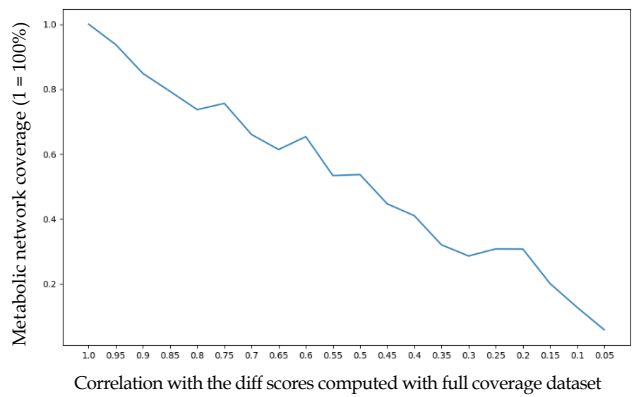

Fig.12. Simulation study results with the baseline approach

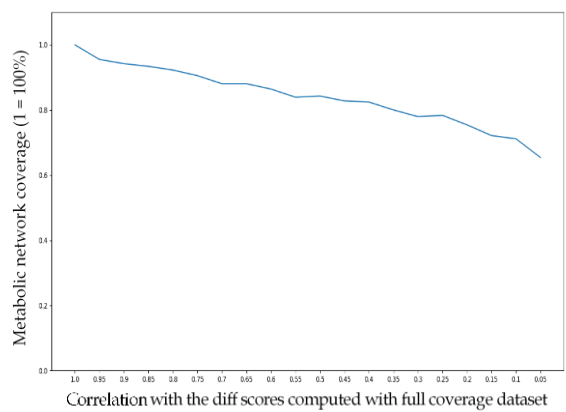

Fig. 13. Simulation study results with the breast cancer dataset employed as a seed

Fig. 12. The correlation decreases proportionally to the coverage reduction, and Metabolitics underperforms in this analysis. This may be attributable to the unrealistic metabolomics data generation approach employed in this baseline run, as it unrealistically assumes that the concentrations of metabolites are independent from each other.

Next, as a more realistic approach, we generated a new metabolomics data set by using the original breast cancer data set as a seed, and expanded it into a full coverage metabolomics data set as follows: For each unmeasured metabolite m, within the seed metabolomics data, we identified the metabolite that is most similar to m in terms of the reactions that it participates. That is, given two metabolites, we use the number of common reactions that they both participate as a proxy to their similarity. Assume that the most similar metabolite for m is x. Then, we assign the fold change of m as that of x plus a certain noise which is generated randomly with a normal distribution with a mean of 0 and a standard deviation of 0.1. Then, the same procedure as in the previous simulation is repeated. Results are shown in Fig. 13. The figure shows that even with very sparse measurements, Metabolitics can provide highly correlated results with full coverage.

## 4 DISCUSSION

Through the analysis of a breast cancer, Crohn's disease, and colorectal cancer metabolomics data, we show that Metabolitics makes biologically relevant discoveries as supported by the literature. The assigned diff scores are consistent with the currently known understanding of breast cancer. What is more, we also discovered two breast cancer associated pathways, namely, Butanoate metabolism (significant decrease in activity), Arginine and Proline Metabolism (significant increase in activity), which have not been reported before in the literature.

Besides, Metabolitics-assigned diff scores allow to distinguish breast cancer patients from healthy people with high f1-score (90%) in a non-invasive and privacy-friendly way based on only blood test samples. Such high classification performance would greatly shrink the need for more advanced and screening procedures (e.g., mammograms), thereby reducing the medical costs on patients and governments. Offering convenient and highly accurate non-invasive diagnosis tools with the help of Metabolitics may greatly increase the number and frequency of screenings, as well as the rate of early diagnosis so that many lives may be saved before it gets late for treatment or intervention.

In addition, Metabolitics is able to quantify the changes in pathways with no or small number of measurements included in the metabolomics data set. This aspect makes Metabolitics superior than the state of the art (e.g., Pathifier, Paradigm, etc.), as it leads to a larger coverage of the metabolic network. This strength of Metabolitics makes it robust to missing some of the key metabolites in the input metabolomics data set, as demonstrated in our worst-case scenario-based robustness evaluation tests. The ability to provide insights even on the unmeasured parts of the metabolic network makes Metabolitics an invaluable tool for biochemistry, medicine, drug researchers in designing new therapies.

What is more, Metabolitics is not specific to metabolomics data. It may be employed for gene expression data analysis with a small modification. More specifically, the objective function of the dynamically constructed linear programming model would be set based on reactions (through gene→enzyme) relationship, rather than on metabolites. The rest of the workflow will be the same. Given the huge amount of available gene expression data, Metabolitics may enable researchers to make high impact discoveries regarding different diseases.

The major future impact of Metabolitics lies in the medical informatics domain. As an example, Metabolitics may be incorporated into novel medical computing tools that may be used to help physicians make treatment decisions in a personalized manner. For each breast cancer patient, a different existing enzyme inhibitor drug may work. The key point here is to find those vital enzymes that will render the above discovered metabolic mechanisms of a cancer cell non-viable. Once those key enzymes are identified, intervening with their known inhibitor drugs (if exist) may be a treatment option to be considered by an oncologist along with other options. We will develop such a recommender tool as follows: Once metabolic mechanism models are identified by Metabolitics, we will choose enzymes that are highly connected at high-flux regions of the network. We will remove those enzymes one by one, and apply Metabolitics on the remaining metabolic network to see if any feasible solution will be produced. Those enzymes, removal of which makes the metabolic network profile more similar to healthy people than people with the disease will be suggested as candidates for intervention with inhibitor drugs. For enzyme-inhibitor drug information, publicly available sources, e.g. DrugBank [32], may be employed. There are already efforts in this direction (e.g., [70]), and we plan to extend these approaches by incorporating the pre-built machine learning models into the analysis.

Moreover, with a similar approach to the above-described one, Metabolitics may be used to develop in-silico experimental models for drug design. That is, each enzyme (or a set of enzymes) may be considered as a possible drug target candidate. Then, Metabolitics is run with the constraint that the targeted region of the network will have no or minimal flux. Next, the analysis results are compared to the original analysis results (before perturbation). If the new analysis results are classified as belonging to a healthy person by the pre-built machine learning model(s), then the targeted enzyme(s) may be considered as a promising drug candidate to be focused in clinical studies.

Last but not the least, the proposed approach may also be employed in "intra-patient" studies. That is, metabolite concentrations from healthy and tumor tissues may be analyzed and compared to reveal the metabolic differentiation of tissues from the same person. This way, person-to-person metabolic differences may be eliminated from the consideration, and health vs. disease state differentiation may be studied with minimal noise.

We note the following limitations of our study. First, the provided diff values may not be directly validated with wet-lab studies, as it is very difficult to quantify the actual flux change in eukaryotic cells in vivo. Second, we associate blood-level measurements to the metabolic network chances in patient cells. However, the measured metabolite changes may be also resulting from the body's response to the existence of a tumor in an individual. Hence, the reported metabolic changes may not be directly associated with the tumor cells. Thus, we use a more general scope as "patient cells" which may belong to healthy or tumor tissues. Nevertheless, Ganti et al. [16] compared simultaneous metabolomics measurements from tumor tissue, blood, and urine, and show that blood is a "decent proxy for the tumor metabolism". Many others (e.g., [21], [37], etc.) also suggested that blood metabolite levels are associated with that of tumor tissues. Finally, the coverage of the human metabolic network by the measured metabolites is still low, which leads to large upper-lower flux boundary intervals possibly obscuring some of the metabolic changes taking place in patient cells.

## 5 Conclusion

Metabolic networks in patients often experience dramatic perturbations in comparison to healthy individuals. These perturbations carry important information regarding the root cause of the underlying condition. Hence, characterizing such changes is invaluable to (i) accurately diagnose patients, and (ii) prescribe a personalized treatment strategy. In this paper, we present Metabolitics algorithm. Given some metabolomics measurements of an individual and a database containing metabolic network data, Metabolitics computes the metabolic network configuration that may lead to the measured changes in the provided metabolomics data. More specifically, it computes a personalized "diff" score for each pathway that represents the amount and direction of activity change in the pathway of that individual. Our approach is based on dynamically creating a linear programming model of the metabolic network according to the given metabolite measurements, and performing a customized flux variability analysis on these models to quantify the differentiation of pathways in reference to healthy people. After computing diff scores, we built classification models which employ the computed diff scores as features to differentiate patients from healthy individuals. We extensively evaluated Metabolitics on three different disease datasets, namely, breast cancer, Crohn's disease, colorectal cancer, and show that it can identify patients with around 90% accuracy (f1-score), and it provides biologically relevant metabolic analysis of diseases.